\documentstyle[aps,pra,twocolumn]{revtex} 
\newcommand{\kbar}{k\hspace{-0.45em}\raisebox{0.7ex}{-}\hspace{0.2em}} 
\def\be{\begin{eqnarray}} 
\def\ee{\end{eqnarray}}

\def\ket#1{| #1 \rangle} 
\def\bra#1{\langle #1 |}

\draft 
\begin{document} 
\preprint{Version 1.0}
\title{Control of dynamical localization by an additional quantum degree of freedom} 
\author{K.\ Riedel, P.\ T\"orm\"a, V.\ Savichev, and W.\ P.\ Schleich}
\address{
Abteilung f\"ur Quantenphysik, Universit\"at Ulm, D-89069 Ulm, Germany} 
\date{\today} 
\maketitle 
 
\begin{abstract} 
We identify a new parameter that controls the localization length 
in a driven quantum system.  
This parameter results from an additional quantum degree of freedom. 
The center-of-mass motion 
of a two-level ion stored in a Paul trap and interacting with  
a standing wave laser field exhibits this phenomenon. 
We also discuss  the influence of spontaneous 
emission.  
\end{abstract} 

\section{Introduction} 
The phenomenon of localization manifests itself in many 
quantum mechanical systems ranging from the localization of light in 
a random medium \cite{lagendijk}, to   Anderson localization of an electronic wave 
\cite{Anderson} and to atoms 
in time-dependent laser fields \cite{review,Moore}. In all these cases the underlying 
classical system is chaotic and shows diffusion as a function of 
time. In contrast, the quantum mechanical counterpart has a localized 
wave function whose width is governed by the classical diffusion and 
Planck's constant \cite{Haake92,Casati95}. In the present paper we show that there exists 
an additional quantum parameter that controls the localization length. 
In  the system of a two-level ion stored in a Paul trap \cite{Paul} and interacting  
with a standing wave it is the detuning between the transition frequency of the ion 
and the laser frequency.  
 
A recent paper \cite{Elghafar97} has shown that a stored ion  
moving along a far-detuned standing wave shows 
localization in position and momentum variables.  In this work we 
have neglected the internal structure of the ion. We have therefore 
only considered the quantum dynamics of a particle moving in a 
one-dimensional  time-dependent potential. In the present 
paper we extend this analysis   
and consider the motion of an ion taking into account
its internal dynamics. 

Our paper is organized as follows: In Sec.\ \ref{2} we first summarize
the essential ingredients of the problem. In particular, we introduce
the relevant equations and describe the methods of solution 
for the classical and quantum mechanical equations of motion.
We also define the quantities which we calculate such as the
position and momentum distributions and their moments.

In Sec.\ \ref{3} we discuss the dependence of the classical and 
quantum mechanical position and momentum distributions on the detuning.
We find characteristic oscillations in the widths of these
distributions  as a function of the detuning.
These oscillations are absent in the corresponding classical curves.
We explain these structures by transforming the Hamiltonian
into an interaction picture. 

We emphasize that this
is different to  the case of 
atoms moving in a phase modulated standing wave. There  the
corresponding  oscillations  appear  in the widths of the quantum as well 
as of the
classical momentum distributions \cite{Robinson95,Bardroff95,Graham96}.
Hence, the detuning is an additional parameter controlling the width
and the shape of the quantum distributions.

Moreover, in the quantum systems 
discussed so far in the context of dynamical localization the quantum 
diffusion is always slower than the classical one.  However, in the
present model the new  control parameter detuning can create situations
in which the quantum diffusion temporarily exceeds the classical one.

Quantum interference effects such as dynamical localization
are extremely sensitive to decoherence arising for example 
from noise  or spontaneous emission \cite{new}.
We are therefore forced to investigate in Sec.\ \ref{4} the influence of spontaneous emission.
We find that the
drastic difference between quantum and classical behavior is
preserved in the limit of  a far off--resonance situation.
We conclude in Sec.\ \ref{5} by summarizing 
our main results.

\section{Formulation of the  problem} 
\label{2}
We consider the standard Paul trap set-up realized experimentally in 
many labs \cite{Wineland,Walther,Toschek}: a standing electromagnetic  
wave of 
frequency $\omega_L$ and wave number  $k$ aligned along the $x$-axis 
couples the internal states $\ket{g}$ and $\ket{e}$ of a single two-level ion of mass $m$ to 
the center-of-mass motion. The resulting dynamics of the ion follows 
from the time-dependent Schr\"odinger equation with the Hamiltonian 
\be 
\hat{\tilde{H}} &=& \frac{\hat{\tilde{p}}^2}{2m}+\frac{1}{2} 
\frac{m\omega^2}{4}\left[a+2q\cos\left(\omega \tilde{t}\right)\right] 
\hat{\tilde{x}}^2 +\frac{1}{2}\hbar\omega_0 \hat{\sigma}_z \nonumber \\ 
&+& \hbar\tilde{\Omega}_0 \hat{\sigma}_x \cos(k\hat{\tilde{x}}) 
\cos(\omega_L \tilde{t})  .   
\label{Schr} 
\ee 
Here the parameters $a$ and $q$ are proportional  
\cite{Paul} to the DC and AC voltages applied to the trap
and $\sigma_x, \sigma_z$ are the standard Pauli matrices. 
Moreover, we denote the frequencies 
of the AC field,  the atomic transition,  the  Rabi frequency  
by $\omega, \omega_0$, and $\tilde{\Omega}_0$, respectively.  
 
We introduce the dimensionless  position  
$\hat{x}\equiv k \hat{\tilde{x}}$, time $t\equiv \omega \tilde{t}/2$  
and momentum $\hat{p}\equiv\frac{2 k}{m \omega} \hat{\tilde{p}} $.  
When we transform into an interaction picture with the unitary
transformation $\mbox{exp}(i \omega_L \sigma_z t/2)$ the dimensionless
Hamiltonian  in rotating wave approximation reads
\begin{eqnarray}  
\hat{H} &\equiv& \frac{4k^2}{m\omega^2} \hat{\tilde{H}} =  
\frac{1}{2} \hat{p}^2+\frac{1}{2}\left[ a+2q 
\cos ( 2 t) \right] \hat{x}^2 - \kbar \Delta \hat{\sigma}_z \nonumber \\ 
&+& \kbar \Omega_0 \hat{\sigma}_x \cos \hat{x}. \label{ham} 
\end{eqnarray} 
Here we have defined the dimensionless Rabi frequency  
$\Omega_0 \equiv \tilde{\Omega}_0/ 
\omega$ and detuning $\Delta = (\omega_L - \omega_0)/\omega$.
 
The dynamics of the ion follows from  the time--dependent 
Schr\"odinger equation 
\begin{eqnarray} 
i \kbar \frac{\partial |\Psi (t)\rangle}{\partial t}  
= \hat{H} |\Psi (t) \rangle 
\end{eqnarray} 
for the state vector 
\begin{eqnarray} 
|\Psi (t)\rangle = \int dx [\Psi_g (x,t) |g\rangle + 
\Psi_e (x,t) |e\rangle ] |x\rangle .  \label{t_ev}  
\end{eqnarray} 
Here the effective Planck constant $\kbar\equiv
2 k^2 \hbar/(m \omega)$  
is consistent with the commutation relation $[\hat{x},\hat{p}] =  
2 k^2/(m \omega)[\hat{\tilde{x}},\hat{\tilde{p}}] = 
2 k^2/(m \omega) i \hbar \equiv i \kbar$.  

The state vector $|\Psi \rangle$ provides the probabilities
\be
P_g(x,t) \equiv |\Psi_g(x,t)|^2
\ee
or
\be
 P_e(x,t) \equiv |\Psi_e(x,t)|^2
\ee
to find the ion at the time $t$ at position
$x$ given it is in the internal state $|g\rangle$ or $|e\rangle$, respectively.
When we are not interested in the internal states the position probability
distribution $P(x,t)$ reads
\be
P(x,t) \equiv |\Psi_g(x,t)|^2 + |\Psi_e(x,t)|^2 .
\ee
 
We simulate the effect of spontaneous emission by the quantum  
Monte--Carlo method \cite{montec,carmichael} 
using the effective non--hermitian Hamiltonian 
\begin{displaymath} 
\hat{H}_{\mbox{eff}} \equiv \hat{H} - i\kbar \frac{\gamma}{2} \hat{\sigma}_+  
\hat{\sigma}_-. 
\end{displaymath} 
Here we introduced the atomic operators $\sigma_+ \equiv |e\rangle\langle g|$
and $\sigma_- \equiv |g\rangle\langle e|$, and 
$\gamma$ is the spontaneous decay rate scaled by $\omega$. 

The moments of time when a spontaneous emission 
event takes place are chosen 
at random. Then the wave function is projected onto the ground state 
and renormalized, that is
$\ket{\Psi} \rightarrow \hat{\sigma}_- \ket{\Psi}/\langle  
\Psi | \Psi \rangle$. 
The recoil $p \in [-\kbar,\kbar]$ is 
chosen randomly  according to the probability 
distribution \cite{stenholm} 
\begin{displaymath}
N(p) \equiv \frac{3}{8 \kbar}  
\left( 1 + \left(\frac{p}{\kbar}\right) ^2  
\right), 
\end{displaymath}
of dipole radiation.
When the results of single runs are averaged, we obtain 
the same result as predicted by a master equation \cite{montec,carmichael}. 
 
As an initial condition for the internal states we use the superposition 
$(|g\rangle + |e\rangle) / \sqrt{2}$  
and the  the center-of-mass  
wave function is a 
Gaussian of width $\Delta x = \sqrt{\kbar}$.  In order to investigate 
localization we start with this wave packet at
the  origin where the classical 
phase space is a stochastic sea \cite{Elghafar97,classical}. 
We calculate the time evolution using the 
split-operator method \cite{10} with a grid of 8192 points. We control 
numerical errors using an adaptive time step--size algorithm \cite{dipl}. 
 
This numerical integration allows us to find the time dependence
of the position probability distributions $P_g(x,t), P_e(x,t)$
and $P(x,t)$. 
Moreover, we calculate the corresponding momentum distributions 
where the wave functions $\Psi_g$ and $\Psi_e$ in position space 
are replaced by the wave functions in momentum space.  
 
We compare and contrast these quantum results to the dynamics \cite{Zaslavsky85} resulting from 
the classical equations of motion \cite{classical,blochgl}
\be 
\dot{x} &=& p  \nonumber  \\ 
\dot{p} &=& -(a+2q\cos (2t)) x + \kbar \Omega_0 \sin (x) r_1  \label{bl1} 
\ee 
for the center-of-mass motion. These equations are driven by the  
Bloch equations \cite{mandel} 
\be 
\dot{r}_1 &=& - 2 \Delta r_2 \nonumber  \\ 
\dot{r}_2 &=& 2 \Delta r_1 + 2 \Omega_0 \cos (x) r_3 \label{bl2} \\ 
\dot{r}_3 &=& - 2 \Omega_0 \cos (x) r_2  \nonumber 
\ee 
describing the internal dynamics.
Here $r_1$, $r_2$, and $r_3$ denote the in-- and out--of--phase quadratures
of the dipole moment,  and the atomic inversion, respectively.

For the comparison with the quantum mechanical 
results we calculate 4096 trajectories starting from a classical Gaussian  
ensemble centered initially at the origin and having the same widths in  
position and momentum as the quantum wave packet. In this way we obtain 
the classical phase space distribution 
from which we find by integration over $x$ or $p$ the
classical momentum or position distributions
$P_{cl}(p)$ or $P_{cl}(x)$. 

\section{Influence of the detuning}
\label{3} 
In the present section we study the influence of the detuning
of the laser field with respect to the atomic transition
on the dynamics of the system. Here we consider the following
quantities of interest: the classical and quantum distributions,
and their width as a function of time and detuning.

In Fig.\ 1 we show the influence of the internal  
structure of the ion on the 
classical and on the quantum diffusion. Here we 
analyze for various detunings $\Delta$ the time dependence of the widths 
$\Delta x$ (left column) of the classical (thin lines) and quantum  
mechanical (thick lines) position distributions $P_{cl}(x,t)$ and $P(x,t)$ 
(right column).  
Whereas the left column   
emphasizes the time dependence of the widths, 
the right column shows  position 
distributions  averaged over the time interval 
$[450\pi,500\pi]$.   
In order to bring out most clearly the influence of the
detuning
we have  neglected spontaneous emission in this figure.
 
For a  very small detuning  the quantum width $\Delta x$ lies 
well below the classical one.  The latter
increases as a function of time. In contrast, the  quantum 
result displays oscillations around a steady--state value. This 
suppression of the classical diffusion is due to dynamical localization 
as discussed in Ref.\cite{Elghafar97}.   
 
However, for slightly larger 
detunings the quantum curve is still oscillating but reaches partially above the 
classical one. For even larger detunings the quantum curve falls again 
below the classical one. Eventually for  larger detunings it again 
goes partially above. Hence, for a fixed detuning, there exist time 
regimes in which the quantum diffusion is stronger than the classical 
one.  
 
This detuning dependence of the quantum diffusion manifests itself in the shape of 
the quantum mechanical  position distribution. The latter  consists of a 
sharp peak and a broad background. The detuning essentially controls the 
shape of the background: Depending on $\Delta$ we  find
distributions either of 
negative or positive curvature. In contrast, the detuning hardly 
influences the classical distributions shown on the right column by  
thin curves. We emphasize that the momentum distributions not shown here
display almost identical behavior.
 
To discuss the influence of the detuning on the localization length 
we show in Fig.\ 2 the widths $\Delta x$ of the classical and the 
quantum distributions averaged from $t=200\pi$ to $500\pi$.  We note 
characteristic oscillations in the quantum mechanical curve as  
a function of the detuning $\Delta$. These oscillations do not appear 
in the classical curve which decreases monotonously for increasing 
$\Delta$. This feature becomes clear when we recall that the  
width of the classical distribution at a given time is determined by 
the classical diffusion constant. The latter  is proportional to the 
perturbing potential caused by the laser field.  In the limiting case 
of large detuning, the effective potential becomes \cite{Kazantsev90} 
proportional to the effective coupling constant $\Omega_0^2/\Delta$. 
As $\Delta$ increases, the perturbation decreases and hence the 
diffusion is slower.

An interesting domain is a   small region around  
$\Delta =0 $. Here the classical diffusion rate sharply decreases.  
The explanation of this feature follows  from the set of Eqs.  
(\ref{bl1}) and (\ref{bl2}): The  effective  potential provided by
the  laser  
field in Eq.~(\ref{bl1}) is proportional to the factor $r_1(t)$. This
quantity can take  
the maximum value of unity.    
For the resonant case, $\Delta=0$, we have $\dot{r}_1=0$ and hence  
$r_1$ is independent of time and determined by the initial condition.  
The  Bloch vector coordinates corresponding to the initial state  
$(|g\rangle + |e\rangle)/\sqrt{2}$   are $r_1(0)=1, r_2(0)=0$ and $ r_3(0)=0$.  
Hence  $ r_1= 1$ results in  the maximum driving strength of the  
center-of-mass motion.

The curve for the quantum widths shows a distinctly different behaviour 
with a lot of structure. In particular, for $\Delta < 1$ we note 
a well--developed resonance structure and 
two sharp minima and maxima. The minimum 
around $\Delta =0$ is related to the separable quantum dynamics in  
the two diabatic potentials
$V^{(\pm)} = \frac{1}{2} \left[ a+2q \cos(2t) \right] x^2 \pm \hbar \Omega_0 \cos x$. 

To bring this out  most clearly we transform  
from the $\ket{g},\ket{e}$ basis states to their superpositions  
$ \ket{+} \equiv (\ket{g} + \ket{e})/\sqrt{2}$, and   
$ \ket{-} \equiv (\ket{g} - \ket{e})/\sqrt{2}$. 
In this basis the Hamiltonian Eq. (\ref{ham})  reads  
\be 
\hat{H} & = &   
\frac{1}{2} \hat{p}^2+\frac{1}{2}\left[ a+2q 
\cos ( 2 t)\right] \hat{x}^2  \nonumber \\  
& & +  \kbar  \Omega_0 \hat{\sigma}_z \cos \hat{x}   +  
  \kbar \Delta \hat{\sigma}_x,  \label{ham_tr}  
\ee  
and the coupling is now  proportional to the detuning $\Delta$.   
Hence we indeed find the potentials $V^{(\pm)}$.

When we now increase the detuning, 
transitions  between the  states  $\ket{+}$ and $\ket{-}$  contribute to the  
quantum diffusion. Therefore, the quantum diffusion rate quickly rises to the 
maximal classical one; this corresponds to the formation of the first 
maximum. 

The next minimum fits  the condition 
$\Delta \approx \omega_s$, where $\omega_s = 0.29$ is the secular 
frequency of the trap.  
The origin of this resonance becomes  clear, when we 
write the Hamiltonian, Eq.\ (\ref{ham}), in 
the interaction picture where the dynamics of the trap as well as the 
internal state energy are transformed away by unitary transformations 
\cite{Bardroff96}.  In this picture the Hamiltonian reads  
\be 
H_{\mbox{int}}(t) = \sum_{n=0}^{\infty} \sum_{k=-[n/2]}^{\infty} 
\sum_{l=-\infty}^{\infty} \kbar && \omega_l^{(n,n+2k)}  
e^{2i\left( l- {k\omega_s}+\Delta\right)t} \times \nonumber \\ 
& &  \sigma_+ \ket{n} \bra{n+2k}     +   \mbox{h.c.}  
\label{hamflo}
\ee 
with the Lamb-Dicke parameter  
$\eta \equiv\displaystyle{ \left[\kbar/(2 \omega_r)\right]^{1/2}} $ 
and 
\be 
&& \omega_l^{(n,n+2k)}  \equiv  \Omega_0 \sqrt{\frac{n!}{(n+2k)!}} [i\eta ]^{2k} 
 \frac{1}{\pi} \times \nonumber  \\ 
& &  \int_{-\pi/2}^{\pi/2} dt [\phi^*(t)]^{2k} 
e^{-\frac{1}{2}\eta^2|\phi(t)|^2} L^{2k}_n(\eta^2|\phi(t)|^2) e^{-2ilt}. 
\ee 
Here $\ket{n}$ denotes the $n$--th energy eigenstate of the time 
independent reference oscillator \cite{Glauber92} with frequency $\omega_r$. 
Moreover, we have made use of
the Floquet solution $\epsilon(t) = \exp (i\omega_s t) \phi (t)$, 
where $\phi (t) = \phi (t+\pi)$ is a periodic function. The solution  
$\epsilon(t)$ obeys \cite{Glauber92} the differential equation 
\be 
\ddot{\epsilon}+ \left[ a+2q\cos( 2t)\right] \epsilon &=& 0 
\label{mathier} 
\ee 
with $\epsilon(0)=1$ and $\dot{\epsilon}(0)=i \omega_r$.  
 
From the term $e^{2i\left(l-{k \omega_s}+\Delta\right) t}$  
in the Hamiltonian (\ref{hamflo}) we  expect 
resonance effects, when $\Delta$ satisfies the condition $l-{k \omega_s}+\Delta = 0$. 
For $l=0, k=1$ we have $\Delta=\omega_s$, which corresponds to   
two--phonon transitions. This resonance suggests an enhancement 
of the quantum diffusion rate on resonance. Instead, we have a deep minimum. 

In order to 
understand this counter-intuitive behavior we calculate  
the characteristic frequencies $\omega_l^{(n,n+2k)}$ 
for $l=0$ and $k=1,2$ at different vibrational quantum numbers $n$. 
From the inset of Fig.\ 2 we recognize that $|\omega_0^{(n,n+2)}|$ has a deep 
minimum around $n=10$, which explains suppression of the diffusion over 
the vibrational states. As shown by the curve in the inset  for  
$|\omega_0^{(n,n+4)}|$ there is no minimum just a decay. 
This causes the quantum diffusion to take on the classical value,  
which explains the formation of the second maximum \cite{Fano}. 
 
We conclude this section by noting that the phenomenon of
oscillations in the width of distributions    
also appears in another quantum  
system showing classical chaos and quantum localization: an atom moving 
in a phase-modulated standing wave 
\cite{Robinson95,Bardroff95}.  However in this case the oscillations are of 
classical origin and appear both in the classical and quantum 
widths.  

\section{Influence of spontaneous emission}
\label{4} 
We now consider the effect of  
spontaneous emission using the quantum Monte--Carlo 
technique \cite{montec,carmichael}.  The purpose of our investigations is twofold. 
First, we want to show that indeed the phenomena discussed so far are  
quantum interference effects. They are therefore sensitive to decoherence 
such as spontaneous emission. The latter  predominantely occures 
when the laser is tuned 
close to resonance with the atomic transition. Second, we want to show 
that in the far-detuned limit the effect of decoherence is negligible and 
the phenomenon of dynamical localization survives.  
 
In Fig.\ 3 we show the results of our  simulations with a realistic rate 
of spontaneous emission corresponding to the decay 
rate of the $S\rightarrow P$ transition at $19.4$ MHz in $^9\mbox{Be}^+$ 
\cite{Jefferts95}. We compare them to the classical and to the 
quantum result without noise. The loss of the coherence causes destruction of 
localization. In this case the widths of the quantum mechanical 
position and momentum distributions are 
larger than the classical ones. This additional diffusion is caused by 
the random recoil kicks following each spontaneous emission event. 
When we neglect the recoil --- which of course is not realistic 
--- the quantum curve follows the classical one.  The 
position and momentum distributions are of classical type, that is 
on the logarithmic scale they are polynomial 
curves.  Note that the pattern of the standing 
wave appears on top of the position distributions. 
 
In Fig.\ 4 we show the results for $\Delta=1000$, corresponding to a 
detuning of $10$ GHz, a value which was mentioned in 
Ref.\ \cite{Elghafar97}. The small oscillations in the quantum widths $\Delta x$ 
and $\Delta p$ are destroyed, but the main phenomenon, the substantial 
quantum suppression of classical diffusion is still visible.  

It is interesting to note 
that  in a single realization of the dynamics the coherence 
of the two--level superposition is completely destroyed by a single 
spontaneous emission. However, it affects only slightly the motional coherence 
because the population of the excited state is very low. 
Furthermore, the small difference between the results  with and without 
spontaneous emission in Fig.\ 4 is of the same order of magnitude that 
was found for the problem of an atom in
a phase modulated standing wave \cite{Goetsch96}. 
 
In the  position distribution of the ground state, 
there is more probability in the classical-like background than in the 
case of no spontaneous emission.  We have found that this background is 
very slowly growing, and no  real steady state can be 
reached. However, if we increase the detuning, the real 
steady state is  approached asymptotically. Since dynamical 
localization appears for a large range of parameters 
\cite{Elghafar97}, there is a lot of room for optimizing the 
parameters. Thus the phenomenon of dynamical localization in a Paul 
trap could be  observed experimentally. 

In order to observe not only dynamical localization but also the
oscillations in the localization length discussed in the previous
section, one should consider a configuration where decoherence is
weaker even for small detunings, for example a different set of
parameters in the present system, or a different system such as a
Raman transition between two ground states.

\section{Conclusions}
\label{5} 
There are basically two ways of controlling 
dynamical localization: (i) through the classical diffusion and 
(ii) through the quantized nature of the variable showing 
localization. This stands out most clearly in the estimate 
$l\sim D / \kbar^2$ for the localization 
length. Here $D$ is the classical diffusion 
coefficient determined by the perturbing potential. The 
coupling to this potential is called the control parameter, since it 
is a direct way to control the localization length.  The scaled Planck 
constant $\kbar$ describes how important the quantization of the 
system is with respect to the perturbation. 

In this paper we have shown  
that, when the perturbing potential has a quantum character as 
well, the relation $l\sim D / \kbar^2$ is not exactly true anymore; we 
observe oscillations in the localization length which do not appear in 
the classical diffusion rate. The parameter determining the 
oscillations could be called the quantum control parameter. 
 
\acknowledgments 
We thank B.\ Kneer and M.\ El Ghafar for many fruitful discussions. 
P.\ T.\ and V.\ S.\ acknowledge the support of the Deutsche 
Forschungsgemeinschaft. We thank the Rechenzentrum Ulm and the 
Rechenzentrum Karlsruhe for their technical support.

\begin{figure}
\caption{Classical and quantum dynamics of 
the center--of--mass motion of a two--level ion stored
in a Paul trap and interacting with a classical standing light wave.
We show the influence of the
detuning $\Delta $ between the frequencies of the atomic transition 
and the light on the dynamics using
probability distributions $P(x)$
in position averaged over the time period $[ 450\pi, 500\pi ]$ (right column),
and  widths 
$\Delta x \protect \equiv \left[ \langle x^2 \rangle -
\langle x \rangle ^2 \right] ^{1/2} $ 
(left column)
as a function of time. 
For a  very small detuning (top) the quantum mechanical width (thick line) 
lies below
the classical width (thin line). Whereas the width of the classical position
distribution
increases monotonously as a function of time the corresponding
quantum result displays oscillations around
a steady--state value. This suppression of 
the  diffusion is due to  dynamical localization
as discussed in Ref.\ \protect\cite{Elghafar97}.
For slightly larger detunings the
quantum curve is still oscillating but
goes above the classical one. For even larger detunings
the quantum curve falls again below the classical one 
and the  diffusion is suppressed by dynamical localization.
However, for  larger detunings (bottom) the quantum curve
again starts to move across the classical one. We note that
only the quantum curves show a strong dependence on the detuning.
This dependence also manifests itself in the position distributions.
Indeed $\Delta$ influences the shapes of the quantum, 
but not of the classical distributions.
Here and in Fig.\ 2   we do not account for spontaneous
emission. We have used the scaled Rabi frequency $\Omega_0 = 2.24$
of the standing wave and the trap parameters $a=0, q=0.4$ corresponding
to the secular frequency $\omega_s = 0.29$.
The effective Planck constant is $\protect \kbar = 0.29$.}
\end{figure}

\begin{figure}
\caption{Dependence of the classical (dashed line) and quantum mechanical
(solid line) variance $\Delta x$ in position on the detuning $\Delta$. For each value of $\Delta$
we have averaged the  variances over the time window
$[200\pi, 500\pi]$. Whereas the classical curve decays monotonously
with a characteristic maximum on resonance, the quantum curve
shows striking oscillations.
The inset shows the amplitudes $|\omega_0^{(n,n+2)}|$ and $|\omega_0^{(n,n+4)}|$
determining the strengths of the 2--phonon and 4--phonon transitions in their
dependences on the vibrational quantum number $n$. Whereas 
$|\omega_0^{(n,n+2)}|$ has a minimum,
$|\omega_0^{(n,n+4)}|$
decays monotonously.
This leads to a minimum  or a maximum in $\Delta x$
for $\Delta = \omega_s$ or $\Delta =2\omega_s$, respectively.
 The squares mark the values of $\Delta x$
at the detunings used in Fig.\ 1. 
Here and in Fig.\ 3
the parameters are as   
in Fig.\ 1.}
\end{figure}

\begin{figure}
\caption{Influence of spontaneous emission on dynamical localization
for $\Delta =0$, that is on atomic resonance: We show
the  time dependence of the widths $\Delta x$ and $\Delta p$ in
position and momentum (left column) and the  
probability distributions $P(x)$ and $P(p)$ averaged over the time interval 
$[200\pi,250\pi]$ (right column).
Fat, thin and thick curves indicate these quantities in the presence
and  absence of spontaneous emission and for the classical case,
respectively.
We note that spontaneous emission destroys
localization and due to the recoil the quantum widths become even larger
 than the classical ones. In the absence of spontaneous emission
the position distributions  of the ground and excited
state (jagged thin curve) are sharply peaked. Moreover they are 
identical due to the fact, that
the initial internal state is the superposition 
$ (\ket{e} + \ket{g})/ \protect \sqrt{2}$.
Spontaneous emission destroys the localization peak and enhances the population
in the ground (upper fat curve) at the expense of the ions in the
excited state (lower fat curve). In this case the quantum distributions in position and momentum
are even broader than the corresponding classical ones (thick lines).
The pattern of the standing wave reflects itself in 
the position distributions (upper right corner).
It is more pronounced in the quantum mechanical curves than in
the classical one.
The results in the case of spontaneous 
emission have been obtained by averaging over a sufficient number of
runs, in this case 79 runs. 
Here and in Fig.\ 4 the  scaled rate of spontaneous emission is
$\gamma = 2$.}
\end{figure}
 
\begin{figure}
\caption{Influence of spontaneous emission on dynamical localization
for  $\Delta = 1000$,
that is far off--resonance. 
As in Fig.\ 3 we show the widths in position and momentum
as a function of time and the 
corresponding probability distributions averaged over the time   
interval $[200\pi,250\pi]$.
Since we start from the ground state of the ion the strong
detuning prevents the population of the excited state.
We therefore only display the ground state population (fat line). We emphasize
that this  curve differs only slightly from the corresponding curve
in the absence of spontaneous emission (thin line).
Spontaneous emission does not destroy the localization. Due to the
small amount of spontaneous emission an average over 49 runs 
was sufficient. In order to make contact with Ref.\protect\cite{Elghafar97}
we used the parameters $\Omega_0 = 94.69$ and $\protect \kbar = 0.0725$.}
\end{figure} 
 

\begin{references}
\bibitem{blochgl}
In deriving these equations, we have assumed
that the quantum
mechanical position   distribution is very
narrow and localized near the classical trajectory
$x(t)$. This property allows us to take the slowly varying cosine
potential out of the integrals.
\end{references}
\end{document}